# Polaritonic self-hybridized critical coupling by Brillouin zone folding induced quasi-bound states in the continuum in bulk $WS_2$ metasurfaces

Kuidong Wang,[†*] Xiongwei Guo,[†] Fu Liu, Jinhai Si, Xun Hou

*Key Laboratory for Physical Electronics and Devices of the Ministry of Education and Shaanxi Key Lab of Information Photonic Technique, School of Electronic Science and Engineering, Xi'an Jiaotong University, Xi'an, 710049, China*

*Contact author: wangkuidong@xjtu.edu.cn

[†]These authors contributed equally to this work.

## Abstract

Maximizing the absorption of exciton-polaritons in transition-metal dichalcogenides (TMDCs) under critical coupling conditions is crucial for developing high-efficiency polariton-based photodetectors, solar cells, and perfect absorbers. Here, we propose a polaritonic critical self-hybridized coupling between Brillouin zone folding driven quasi-bound states in the continuum (BZF-QBIC) mode and *A* excitons in a bulk $WS_2$ metasurface. This polaritonic critical coupling shows robust behavior with respect to the asymmetric parameters of the involved BZF-QBIC mode, in contrast to the asymmetry-sensitive conventional QBIC-based critical coupling. By changing the height of $WS_2$ metasurfaces, we observe a typical anti-crossing behavior with a Rabi splitting of 246.5 meV. Moreover, the maximum absorption value of 0.5 for exciton-polaritons is achieved at the polaritonic critical coupling condition. The large population of excitons provided by bulk $WS_2$ is essential for achieving polaritonic critical coupling and can further enhance polaritonic absorption. This work enables the realization of robust, high-absorption self-hybridized polaritonic critical coupling in TMDCs, and further expedite their applications in high-absorption low-dimensional polaritonic devices.

# 1.Introduction

Critical coupling between optical cavities and materials gives new properties for the coupled material[1-5]. This leads to various applications in optoelectronics, such as nonlinear frequency conversion[6], enhanced photoluminescence[7], solar cells[8]. In these coupled systems, the leakage rate of optical cavities and the absorption rate of materials must be equal[9,10]. To date, critical coupling has been demonstrated in both the weak- and strong-coupling regimes[8,11-13]. In particular, strong light-matter critical coupling provides a non-perturbative approach to modify the properties of the couple system and offers potential applications in polariton-related perfect absorbers[14], few-photon nonlinearity[15] and photovoltaics[16]. Such unique characteristics arise from their hybrid light–matter nature, which forms polaritons when the energy exchange rate between the cavity mode and the excitons of the material exceeds damping rates of the individual constituents[17-20]. In order to achieve strong critical coupling, both large oscillator strength of the material and high-quality (high-Q) optical mode of the cavity are required.

TMDCs are ideal van der Waals materials for the realization of strong coupling due to their large exciton oscillator strength and stable tightly bound excitons[21,22], which facilitate the study of strong critical coupling in the coupled TMDCs. A typical high-Q optical mode that supported by lossless metasurfaces is symmetry-protected QBIC[23,24], which arises from a broken in-plane symmetry in a normal QBIC mode. With the existence of such modes, the optical fields can be significantly confined inside the nanocavity, and facilitates the strong critical coupling of TMDCs[13,25,26]. However, the current high-Q modes in symmetry-protected QBICs (SP-QBICs) exist only within a small range of the asymmetric parameters of the in-plane inversion symmetry structure. Moreover, tuning the asymmetry parameter of the SP-QBICs mode to make its damping rate equal to that of the excitonic mode inevitably shifts its resonant wavelength. This factor strongly complicates the realization of strong critical coupling by robust high-Q QBIC modes in lossless nanocavity, where both the damping rate (Q-factor) and the resonant wavelength of the QBIC mode must match those of the excitonic mode.

Here, we present a novel self-hybridized strong critical coupling system between BZF-QBIC mode and *A* excitons in bulk $WS_2$ metasurface. Specifically, current SP-QBIC-based polaritonic critical coupling[27,28] is sensitive to asymmetric parameters, whose variation can easily shift the optical resonant wavelength, thereby hindering the realization of robust polaritonic critical coupling. Together with the large population of excitons in bulk $WS_2$ metasurface, the proposed coupled system enables self-hybridized strong critical coupling. We demonstrate that the BZF-QBIC modes can be modulated by utilizing different geometric parameters of bulk $WS_2$ metasurfaces, a typical anti-crossing feature with a Rabi splitting of 246.5 meV for the self-hybridized system is observed. Under the critical coupling condition, we enhance the absorption of exciton-polaritons to the maximum value of 0.5, providing an effective way to develop polariton-based high-absorption devices. The proposed critical-coupling based on BZF-QBICs is not limited to the visible range in $WS_2$ platform, but can be extended to other frequency regimes and excitonic or vibrational systems, offering a general strategy for realizing strong critical coupling. This work introduces a new way to achieve polaritonic self-hybridized critical coupling, facilitating the application of high-efficiency low-dimensional semiconductor polaritonic devices.

## 2. Numerical Models

Figure 1(a) illustrates the schematic of the self-hybridized structure, where the bulk $WS_2$ metasurface is placed onto a $SiO_2$ substrate. We choose bulk $WS_2$ due to its strong $A$ exciton oscillation intensity (compared to $MoS_2$, $MoSe_2$ and $WSe_2$), as well as its high refractive index ($n = 4.2$ at $\lambda = 633$ nm). In this work, the refractive index of $SiO_2$ is set to $n = 1.45$, and the in-plane complex permittivity of bulk $WS_2$ is described by a single Lorentz oscillator model[29]

$$\varepsilon(E) = \varepsilon_B + f_0 \frac{E_{exc}^2}{E_{exc}^2 - E^2 - iE\Gamma_{exc}} \tag{1}$$

where $\varepsilon_B = 18$, $f_0 = 0.4$, $E_{exc} = 1.96$ eV, and $\Gamma_{exc} = 90$ meV represent the background permittivity, the oscillator strength, the resonance energy, and the linewidth of the $A$ exciton, respectively. The effective permittivity of the bulk $WS_2$ metasurface has two contributions, a pure excitonic mode provided by excitons and an optical mode from background-high-index dielectric nanocavity (Figure S1). By choosing a specific parameter for the metasurface, strong coupling between these modes can be reached. For the system only considering the background index without excitonic effects, where oscillator strength $f_0 = 0$. In the case of the pure excitonic contribution, i.e., background permittivity $\varepsilon_B = 1$. Since the interlayer interaction of bulk $WS_2$ has van der Waals nature, the out-of-plane permittivity is set to $7 + 0i$[29]. The calculated in-plane complex permittivity for bulk $WS_2$ are shown in Figures S1(a) and S1(b). In addition, the electromagnetic response of this work is simulated by using the finite element method with COMSOL Multiphysics. The y-polarized plane wave is normally incident on the optical system along the z-axis, and periodic boundary conditions to be applied in the x and y directions, with perfectly matched layers boundary condition is applied in z direction.

# 3. Results and Discussion

## 3.1 The Optical Properties of BZF-QBIC mode

To confirm the existence of BZF-QBIC mode in our proposed bulk $WS_2$

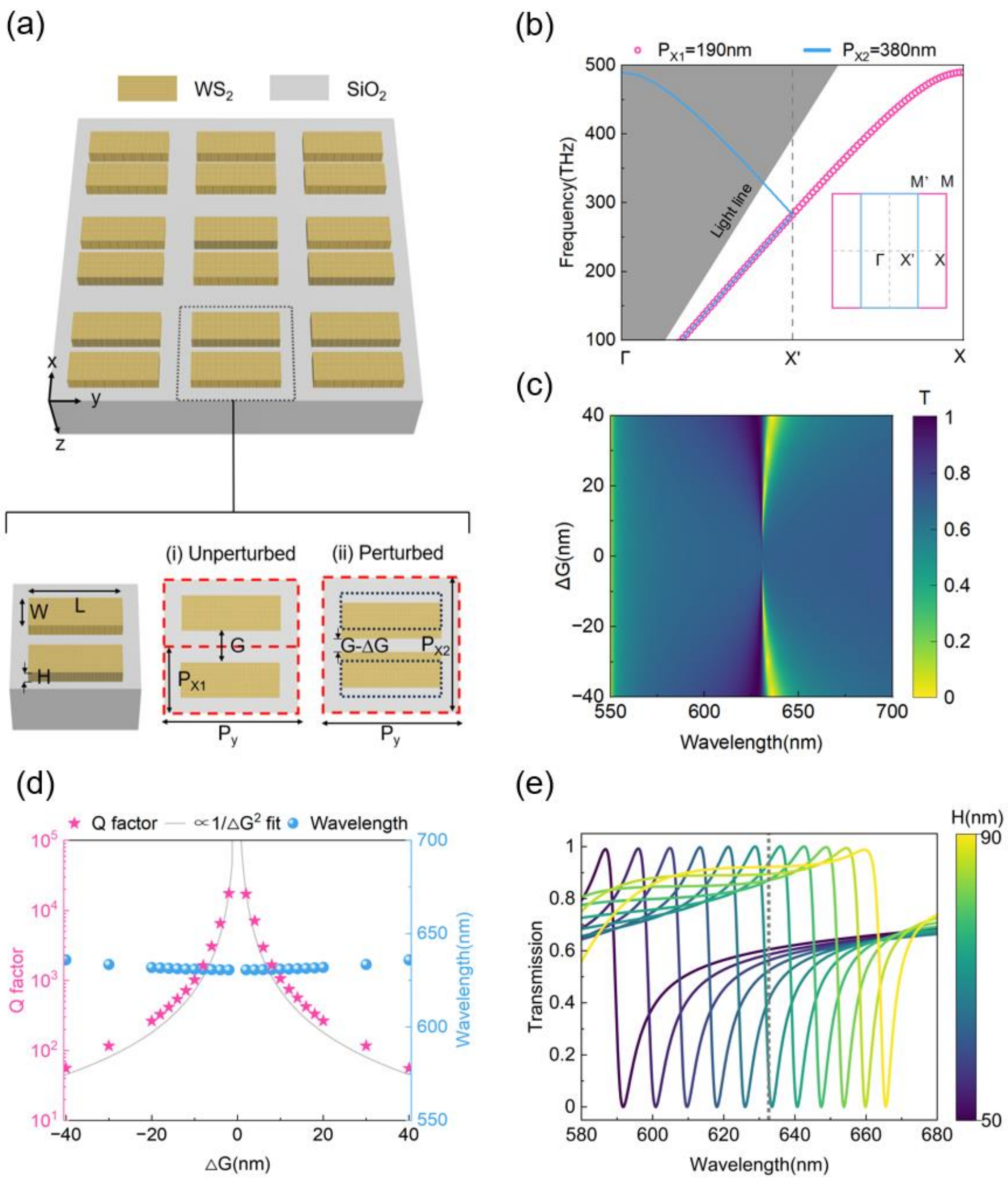

**Figure 1.** Optical properties of BZF-QBIC mode in bulk $WS_2$ metasurfaces. (a) Schematic of self-hybridized bulk $WS_2$ metasurface unit cell: (i) without and (ii) with gap perturbation $\varDelta G$, resulting in a period-doubling in the x-direction. (b) Calculated transverse-magnetic (TM) band structures of unperturbed (hollow pink circles) and perturbed (solid blue curve) metasurfaces, with light line as a reference. (c) Transmission spectrum as a function of gap perturbation $\Delta G$ for $H$ = 70 nm, $L$ = 280 nm and $W$ = 100 nm. (d) Simulated Q factors (pink stars) and wavelength (blue dots) extracted from (c) with respect to gap perturbation $\Delta G$. (e) Transmission spectrum of bulk $WS_2$ metasurface as a function of rod height $H$ with $\Delta G$ = 30 nm, $L$ = 280 nm and $W$ = 100 nm. The black dashed line denotes the excitonic wavelength of $A$ exciton of $WS_2$.

metasurface, we only consider the background-high-index metasurface by setting the oscillator strength $f_0$ to zero. As shown in Figure 1(a), the $WS_2$ nanorods are designed with dimensions of $H = 70$ nm, $L = 280$ nm and $W = 100$ nm on purpose for the realization of hybridized strong coupling between the excitonic resonance and the optical mode of the metasurface discussed below. For the unperturbed structure ($\Delta G = 0$ nm, Figures 1(a)-(i)), the unit cell consists of a single $WS_2$ nanorod ($G = 90$ nm) with periods of $P_{x1} = 190$ nm and $P_y = 380$ nm along the x- and y-directions, respectively. When the gap perturbation is introduced ($\Delta G \neq 0$ nm, Figure 1(a)-(ii)), this only induces a period doubling of the unit cell in the x-direction ($P_{x2} = 380$ nm), leading to the fold of the first Brillouin zone of the bulk $WS_2$ metasurface (the inset of Figure 1(b)). To demonstrate the effect of the gap perturbation, Figure 1(b) compared the transverse-magnetic (TM) band structure of these two cases. For the unperturbed case (symmetric case), the fundamental modes are located below the light line, corresponding to guided mode (GM) that cannot radiate into free space. With the introduction of gap perturbation (asymmetric case), the $X$ point of the GM folds into the $\Gamma$ point, lying above the light line (bringing the GM into the radiation continuum). By considering the Q factor of this folded mode in momentum space (Figure S2), and comparing the Q factor properties with that of GM resonance (flat Q factor)[30,31], this folded mode exhibits a rapid decrease of Q factor away from $\Gamma$ point in momentum space, indicating its BZF-QBIC nature. The origin of the BZF-QBIC can be revealed by multipolar decomposition of the mode in the Cartesian coordinates[32,33] (see Supporting Information for details). As shown in Figure S3(a), the mode is dominated by electric toroidal dipole (ETD), which is consistent with the near-field distributions shown in Figure S3(b).

In order to quantify the properties of BZF-QBIC mode for self-hybridized strong coupling in our case, it is necessary to reveal the $\Delta G$-dependent Q factor and geometric parameter-dependent dispersion of BZF-QBIC mode. Figure 1(c) shows the transmission spectrum as a function of gap perturbation $\Delta G$. In the symmetric case ($\Delta G = 0$ nm), the resonance vanishes. Whereas for the case of asymmetric condition ($\Delta G \neq 0$ nm), the linewidth of the mode gradually broadens as $\Delta G$ increases. This is a signature of BZF-QBIC mode as shown above. Importantly, the resonant wavelength of this mode is nearly unchanged with respect to $\Delta G$. Such an asymmetry-near-independent resonance behavior, compared to the traditional asymmetry-sensitive QBIC mode, provides more tolerance for high-Q mode when fixing the wavelength. To understand the effect of gap perturbation $\Delta G$ on the Q factor of BZF-QBIC mode, the Q factor was extracted by Fano formula fitting of the transmission spectrum[34,35] (Figure S4), and exhibits an inverse quadratic dependence on $\Delta G$, which keeps high values over a wide asymmetric range (Figure 1(d)). In order to optimize the BZF-QBIC mode for strong coupling in the bulk $WS_2$ metasurfaces, we investigate the height dependence of the BZF-QBIC mode. As the nanorod height increases, the BZF-QBIC mode resonance redshifts with nearly unchanged linewidth, and crossing the bulk $WS_2$ *A*-exciton wavelength (Figure 1(e), white dashed line). A similar dependence on the nanorod length is also observed in Figure S5. Such geometric tunability provides a convenient platform for the subsequent investigation of detuned strong light–matter coupling.

## 3.2 Polaritonic critical coupling in bulk $WS_2$ metasurface

To achieve polaritonic critical coupling, strong coupling between excitons and BZF-QBIC mode in bulk $WS_2$ metasurface was investigated. Here, both the background-dielectric and excitonic refractive indices are included in the following calculations. As mentioned above, the parameters of the metasurface are intentionally chosen to provide a BZF-QBIC mode, whose wavelength matches the *A* exciton of bulk $WS_2$. As shown in Figure 2(a), two new resonant peaks appear at 594.2 nm and 673.8 nm in absorption, reflection, and transmission spectra of the self-hybridized coupled system. These resonances are caused by self-hybridized coupling between *A* exciton and BZF-QBIC mode. Clearly, a residual peak between two coupled peaks can also be observed, which originates from the excitonic resonance of uncoupled $A$ excitons. To further investigate the dispersion relation of the self-hybridized system, the height-dependent absorption spectra were calculated. As shown in Figure 2(b), a clear anti-crossing behavior is observed when changing the height of the nanorod. A similar anti-crossing signature can be also found in length-dependent absorption spectra in Figure S6. This is evidence of strong coupling, where the resonances below and above the excitonic wavelength correspond to the upper polariton (UP) and lower polariton (LP), respectively.

To confirm the existence of strong coupling and estimate the corresponding coupling strength, a widely used coupled oscillator model reads[36]

$$\begin{pmatrix} E_{cav} + i\hbar\gamma_{cav} & g \\ g & E_{exc} + i\hbar\gamma_{exc} \end{pmatrix} \begin{pmatrix} \alpha \\ \beta \end{pmatrix} = E_{UP,LP} \begin{pmatrix} \alpha \\ \beta \end{pmatrix} \quad (2)$$

Here, $E_{cav}$ and $E_{exc}$ denote the energies of the uncoupled BZF-QBIC mode and $A$ exciton, respectively, while $\gamma_{cav}$ and $\gamma_{exc}$ represent their half-width at half-maximum (HWHM). The coupling strength between the BZF-QBIC mode and the excitons is given by $g$, while $\alpha$, $\beta$ are the Hopfield coefficients satisfying the normalization condition $|\alpha|^2 + |\beta|^2 = 1$. Here, $|\alpha|^2$ and $|\beta|^2$ represent the relative contributions of the light and matter components to the polaritons, respectively. The energies of the

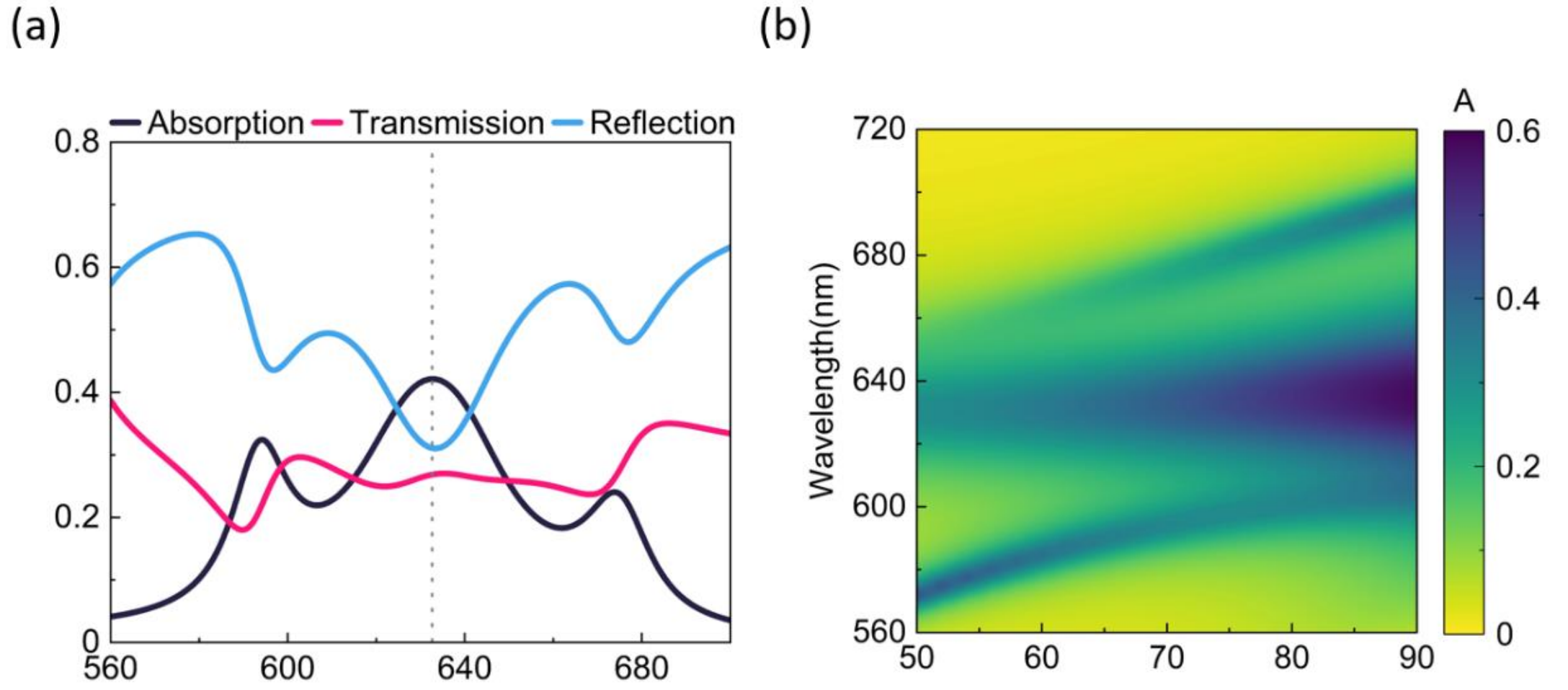

**Figure 2.** Optical properties of self-hybridized strong coupling. (a) The absorption, reflection and transmission spectra of the coupled system. (b) The absorption spectrum of the coupled structure as a function of $WS_2$ nanorods height $H$ with a gap perturbation of 30nm.

UP/LP are represented by $E_{UP,LP}$. Strong coupling is achieved when the following condition is satisfied:

$$\frac{g}{\frac{|\gamma_{cav}-\gamma_{exc}|}{2}} > 1 \tag{3}$$

$$\frac{g}{\sqrt{\frac{\hbar\gamma_{cav}^2+\hbar\gamma_{exc}^2}{2}}} > 1 \tag{4}$$

Relation (3) ensures the existence of the Rabi splitting, while relation (4) guarantees its observability in experiments. When $E_{cav} = E_{exc}$ (zero detuning) is satisfied, a Rabi splitting of 246.5 meV can be extracted from simulated results in Figure 2(b). Considering the dissipation of BZF-QBIC mode and *A* exciton, the coupling strength of the coupled system in our case is estimated to be 124.7 meV. Consequently, self-hybridized system satisfies the criteria (Inequations (3) and (4)) for strong coupling. Such a criteria can also be evidenced via FEM calculations and CMT model (Figure S7).

Based on the self-hybridized strong coupling, the polaritonic critical coupling is an efficient way to enhance the absorption of exciton polaritons. Such a condition can be achieved when $\gamma_{cav} = \gamma_{exc}$. As displayed in Figure 3(a), the damping rate of BZF-QBIC modes and *A* excitons were simulated as a function of gap perturbation $\Delta G$. They hold the same value when $\Delta G = 60$ nm. To study the absorption enhancement behavior of the polaritonic critical coupling, the absorption spectra of the self-hybridized structure with different damping rates: $\gamma_{cav} < \gamma_{exc}$, $\gamma_{cav} = \gamma_{exc}$(critical coupling) and $\gamma_{cav} > \gamma_{exc}$ were simulated under zero detuning ($E_{cav} = E_{exc}$) (Figure 3(b)). In all cases, we observe UP and LP absorption peaks, accompanied by a residual peak arising from uncoupled *A* excitons in bulk $WS_2$. Notably, when $\gamma_{cav} = \gamma_{exc}$, the absorption strength of UP reach the theoretical maximum value of 0.5, which is signature of critical coupling, enhancing polariton absorption through critical strong clearly larger than those under the conditions of $\gamma_{cav} < \gamma_{exc}$ or $\gamma_{cav} > \gamma_{exc}$. This critical coupling is essential, especially for polariton detectors and polariton solar cells, as it can significantly enhance the efficiency of polariton-based devices. Moreover, height-dependent absorption spectra were calculated for the three cases, as shown in Figure S8, where anti-crossing feature is observed in all cases, As indicated in Table 1, a larger linewidth of the BZF-QBIC mode reduces the coupling strength. Furthermore, when $\gamma_{cav} = \gamma_{exc}$, polaritonic critical coupling is achieved (inequation (3), (4) is satisfied). Benefiting from the proposed high-Q BZF-QBIC mode, the exciton polariton self-hybridized critical coupling system presented here exhibits a larger Rabi splitting energy and stronger coupling strength[25], paving the way for a wider range of polaritonic device applications.

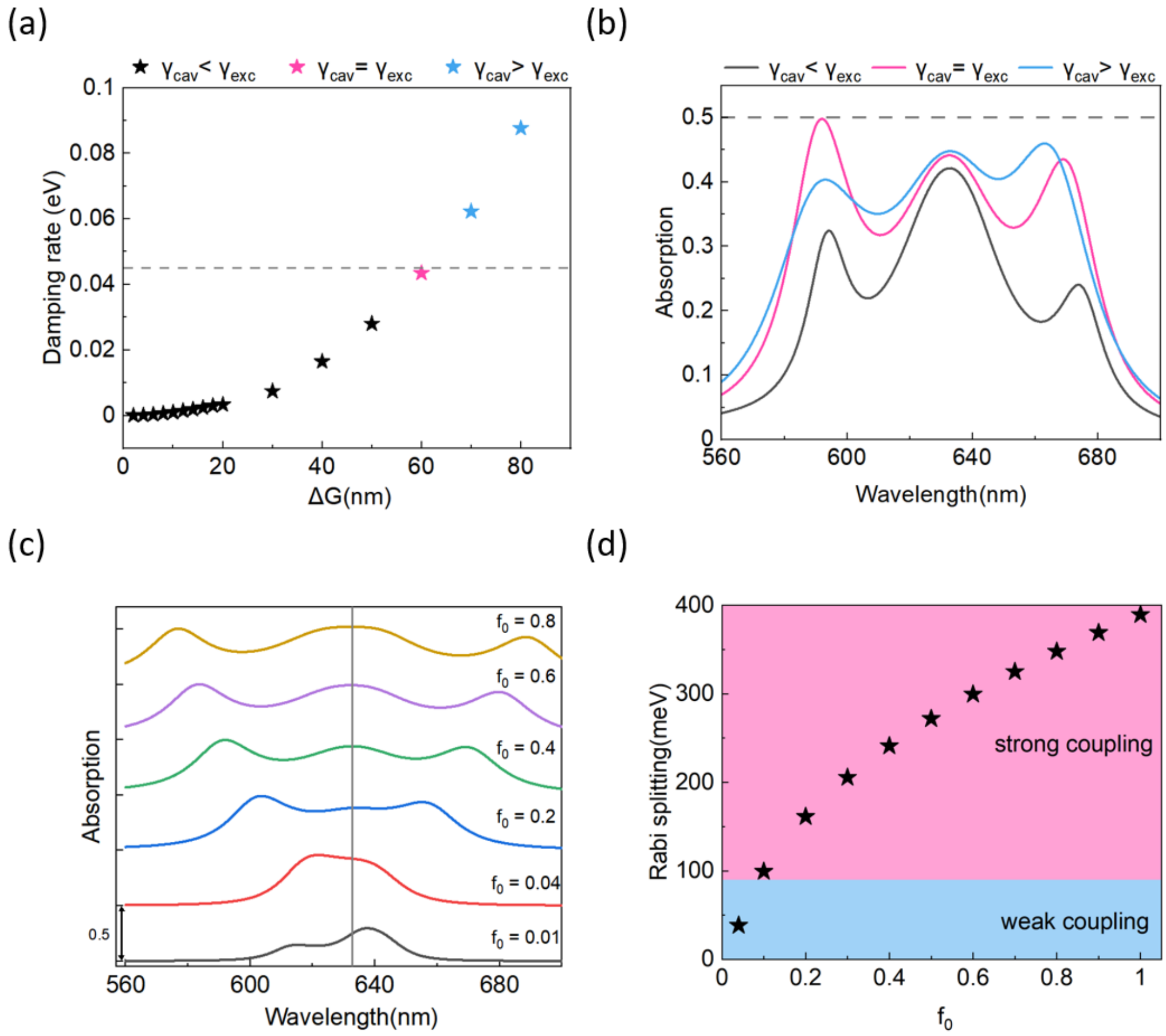

**Figure 3.** Optical properties of polaritonic critical coupling. (a) The damping rate of BZF-QBIC mode $\gamma_{cav}$ and *A* excitons $\gamma_{exc}$ as a functions of gap perturbation $\Delta G$. (b) The absorption spectrum of the self-hybridized metasurfaces at different gap perturbations: 30nm ($\gamma_{cav} < \gamma_{exc}$), 60nm ($\gamma_{cav} = \gamma_{exc}$), 80nm ($\gamma_{cav} > \gamma_{exc}$) under zero detuning. (c) The absorption spectra and (d) Rabi splitting energy of the bulk $WS_2$ metasurfaces as a function of oscillator strength $f_0$ under zero detuning and critical coupling conditions.

To explore the role of bulk WS2 on critical strong coupling, the effect of the population of excitons was investigated in self-hybridized system. The linewidth of the excitonic mode and BZF-QBIC mode are set to be equal to satisfy the critical coupling condition, whereas the population of excitons can be changed by varying the oscillator strength. As shown in Figure 3(c), the absorption spectra were calculated at different oscillator strengths under zero detuning. As the population of excitons decreases, the absorption peak of the uncoupled excitons weakens, accompanied by decreases in both the Rabi splitting energy and coupling strength, driving the self-hybridized system from strong- to weak-coupling regime (Figure 3(d)). Therefore, the large population of excitons provided by bulk $WS_2$ is essential to achieve polaritonic critical coupling. Compared with weak critical coupling, the UP branch in the coupled resonance exhibits stronger absorption under strong critical coupling, consistently reaching the maximum of 0.5. This further demonstrates that a large population of excitons can further enhance the absorption in critical coupling.

This work demonstrates polaritonic critical self-hybridized coupling for a fixed excitonic linewidth. Since the excitonic linewidth in TMDC materials is strongly influenced by temperature or fabrication conditions, employing excitonic materials with narrower linewidths would allow strong critical coupling to be achieved with higher-Q BZF-QBIC modes, thereby enabling larger coupling strengths in self-hybridized polaritonic systems, which could facilitate the development of efficient and robust polariton devices across a broad detuning range.

**Table 1.** The coupling results of the self-hybridized structure for different BZF-QBIC mode damping rate corresponding to Figure S8

| | UP(nm) | LP(nm) | Rabi splitting (meV) | g(meV) | c1 | c2 |
|---|---|---|---|---|---|---|
| $\gamma_{cav} < \gamma_{exc}$ | 594.2 | 673.8 | 246.5 | 124.7 | 6.6 | 3.9 |
| $\gamma_{cav} = \gamma_{exc}$ | 592 | 669 | 241.1 | 120.6 | $\rightarrow +\infty$ | 2.68 |
| $\gamma_{cav} > \gamma_{exc}$ | 593 | 663.1 | 221.1 | 112.6 | 5.3 | 1.6 |

# 4. Conclusions

To conclude, we design a self-hybridized strong critical coupling system between the *A* excitons and the BZF-QBIC mode in bulk $WS_2$ metasurface. The BZF-QBIC mode is particularly suitable for strong coupling due to its asymmetric perturbation-near-independent resonant, along with its high-Q mode. Our results show that the bulk $WS_2$ metasurface is at strong coupling regime with a Rabi splitting of 246.5 meV. In addition, we reveal that strong critical coupling can modulate polariton branch populations, enhancing the absorption of polaritons to a maximum value of 0.5 under single-port input. Our study provides a novel idea for polaritonic self-hybridized critical coupling in bulk TMDCs, paving the way for applications in high-efficiency polaritonic devices.

# 5. Acknowledgements

We acknowledge support from the National Natural Science Foundation of China under Grant No. 12304371 and the Young Talent Support Plan of Xi'an Jiaotong University.

Supporting Information

# Polaritonic critical self-hybridized coupling by Brillouin zone folding induced quasi-bound states in the continuum in bulk $WS_2$ metasurfaces

Kuidong Wang,[†*] Xiongwei Guo,[†] Fu Liu, Jinhai Si, Xun Hou

*Key Laboratory for Physical Electronics and Devices of the Ministry of Education and Shaanxi Key Lab of Information Photonic Technique, School of Electronic Science and Engineering, Xi'an Jiaotong University, Xi'an, 710049, China*

*Contact author: wangkuidong@xjtu.edu.cn

[†]These authors contributed equally to this work.

## 1. Dispersion of bulk $WS_2$ metasurface.

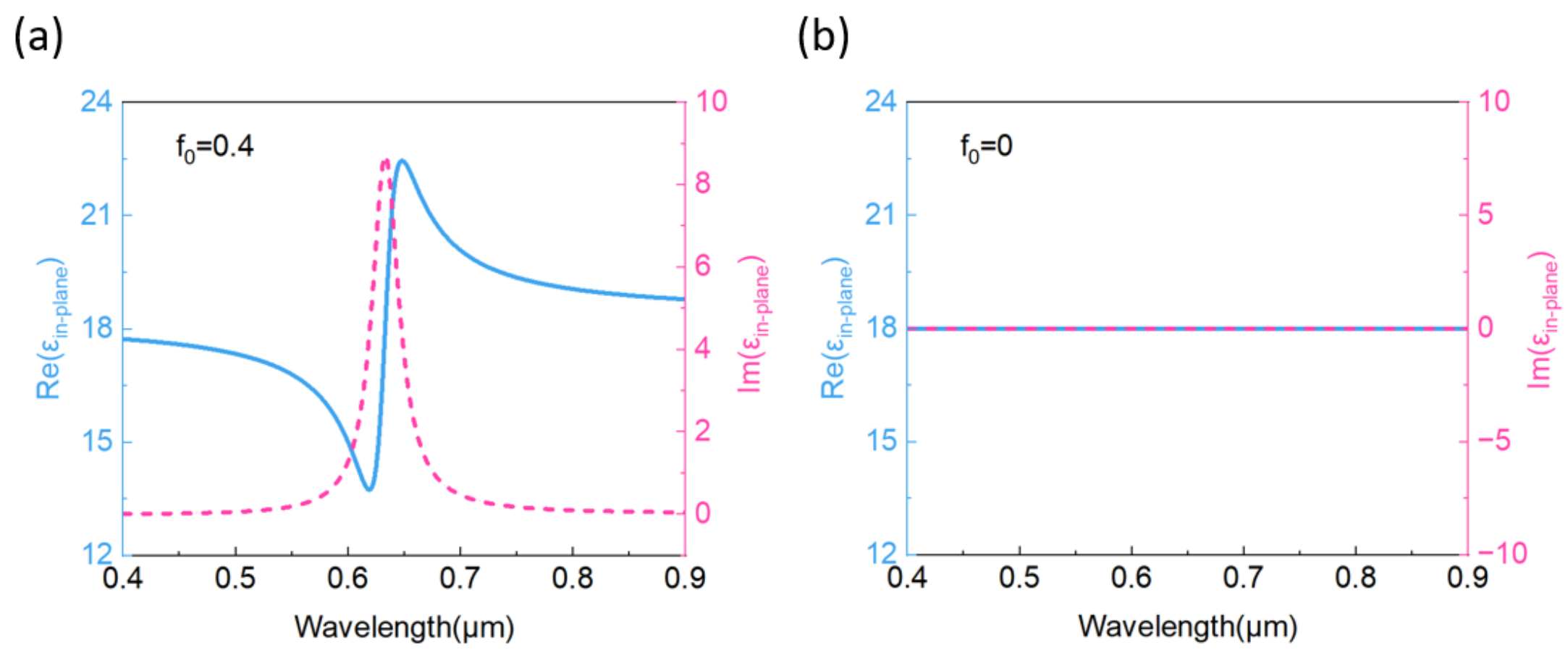

**Figure S1.** The real (solid blue curve) and imaginary (dashed red curve) parts of the permittivity for the bulk $WS_2$ with (a) pure excitonic contribution of $f_0 = 0.4$ and (b) background index cavity of $f_0 = 0$.

## 2. The Q-factor of BZF-QBIC mode in momentum space.

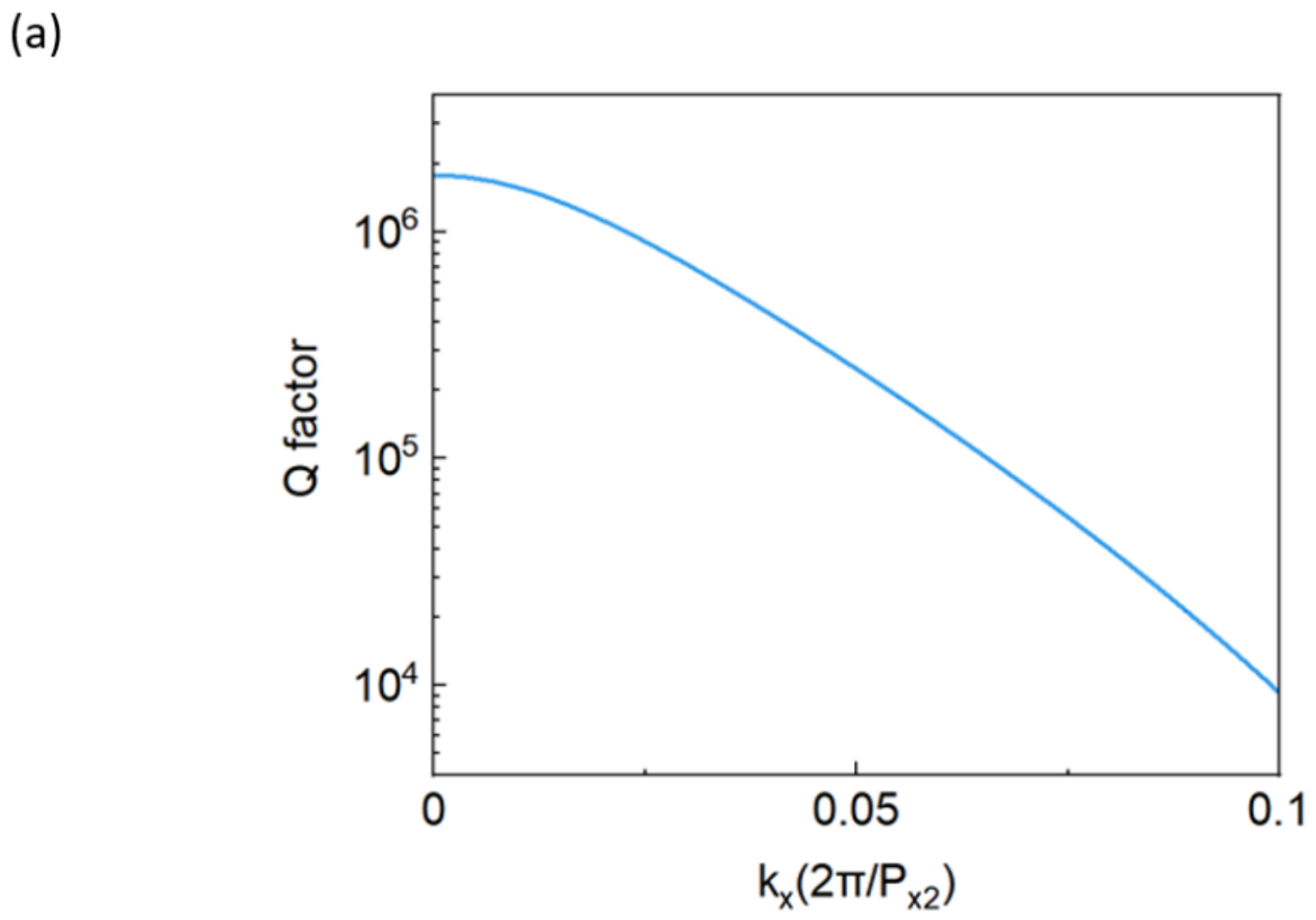

**Figure S2.** Simulated Q-factor of BZF-QBIC mode in momentum space.

## 3. Multipole decomposition and near-field analysis of the BZF-QBIC mode.

We analyzed the far-field radiation of the BZF-QBIC mode by the standard multipole decomposition method, and calculated the contributions from the electric dipole (ED, $P_\alpha$), magnetic dipole (MD, $M_\alpha$), electric toroidal dipole (ETD, $T_\alpha^{(e)}$), magnetic toroidal dipole (MTD, $T_\alpha^{(m)}$), electric quadrupole (EQ, $Q_{\alpha\beta}^{(e)}$), and magnetic quadrupole (MQ, $Q_{\alpha\beta}^{(m)}$). In Cartesian coordinate system ($\alpha, \beta = x, y, z$), ED, MD, ETD, MTD, EQ, and MQ can be expressed as[1,2]

$$P_\alpha = \frac{1}{i\omega}\int \vec{j} d^3r \tag{1}$$

$$M_\alpha = \frac{1}{2c}\int (\vec{r} \times \vec{j}) d^3r \tag{2}$$

$$T_\alpha^{(e)} = \frac{1}{10c}\int [(\vec{r} \cdot \vec{j})\vec{r} - 2r^2\vec{j}] d^3r \tag{3}$$

$$T_\alpha^{(m)} = \frac{k^2}{20}\int r^2(\vec{r} \times \vec{j}) d^3r \tag{4}$$

$$Q_{\alpha\beta}^{(e)} = \frac{1}{2i\omega}\int \left[r_\alpha j_\beta + r_\beta j_\alpha - \frac{2}{3}(\vec{r} \cdot \vec{j})\delta_{\alpha,\beta}\right] d^3r \tag{5}$$

$$Q_{\alpha\beta}^{(m)} = \frac{1}{3c}\int [(\vec{r} \times \vec{j})_\alpha r_\beta + (\vec{r} \times \vec{j})_\beta r_\alpha] d^3r \tag{6}$$

$c, r, \omega, n, k, \vec{E}$ and $\vec{j} = i\omega(n^2 - 1)\vec{E}$ denote the speed of light, the location, the angular frequency of light, the refractive index, the wave vector, the electric field, and the current density, respectively. The scattering powers corresponding to each multipole moment contribution can be given as

$$I_P = \frac{2\omega^4}{3c^3}|P_\alpha|^2 \tag{7}$$

$$I_M = \frac{2\omega^4}{3c^3}|M_\alpha|^2 \tag{8}$$

$$I_{T^{(e)}} = \frac{2\omega^6}{3c^5}|T_\alpha^{(e)}|^2 \tag{9}$$

$$I_{T^{(m)}} = \frac{2\omega^6}{3c^5}\left|T_\alpha^{(m)}\right|^2 \tag{10}$$

$$I_{Q^{(e)}} = \frac{\omega^6}{5c^5}\sum|Q_{\alpha\beta}^{(e)}|^2 \tag{11}$$

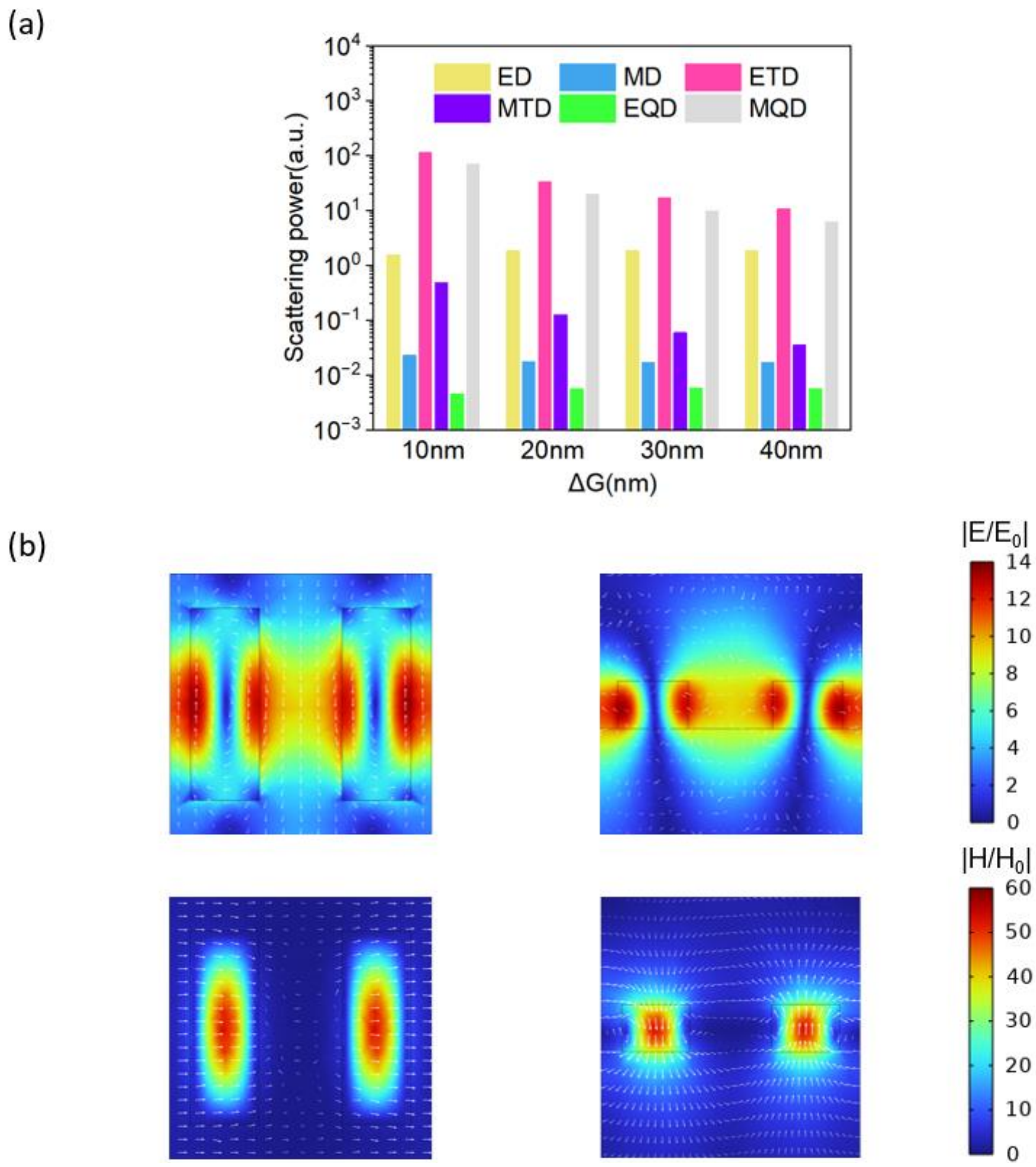

**Figure S3.** (a) The scattering power of different multipole moments with different gap perturbation $\Delta G$. (b) Distribution of electric and magnetic fields for the $WS_2$ metasurface of BZF-QBIC mode with the gap perturbation $\Delta G = 30$nm.

$$I_{Q^{(m)}} = \frac{\omega^6}{40c^5}\sum|Q_{\alpha\beta}^{(m)}|^2 \quad (12)$$

As shown in Figure S3(a) and (b), the BZF-QBIC mode is dominated by ETD,

## 4. Using the Fano formula to fit the transmission spectrum

The simulated transmission spectrum (red star) can be fitted by the Fano formula (gray solid curve) [3,4]

$$T_{Fano}(\omega) = \left|a_1 + ja_2 + \frac{b}{\omega-\omega_0+j\gamma}\right|^2 \quad (13)$$

where $\boldsymbol{a_1}$, $\boldsymbol{a_2}$ and $\boldsymbol{b}$ are constant real numbers, $\boldsymbol{\omega_0}$ is the central resonant frequency. The radiation dissipation $\boldsymbol{\gamma}$ can be obtained from the Fano line shape, from which the Q factor of optical mode in bulk $WS_2$ metasurfaces can be determined.

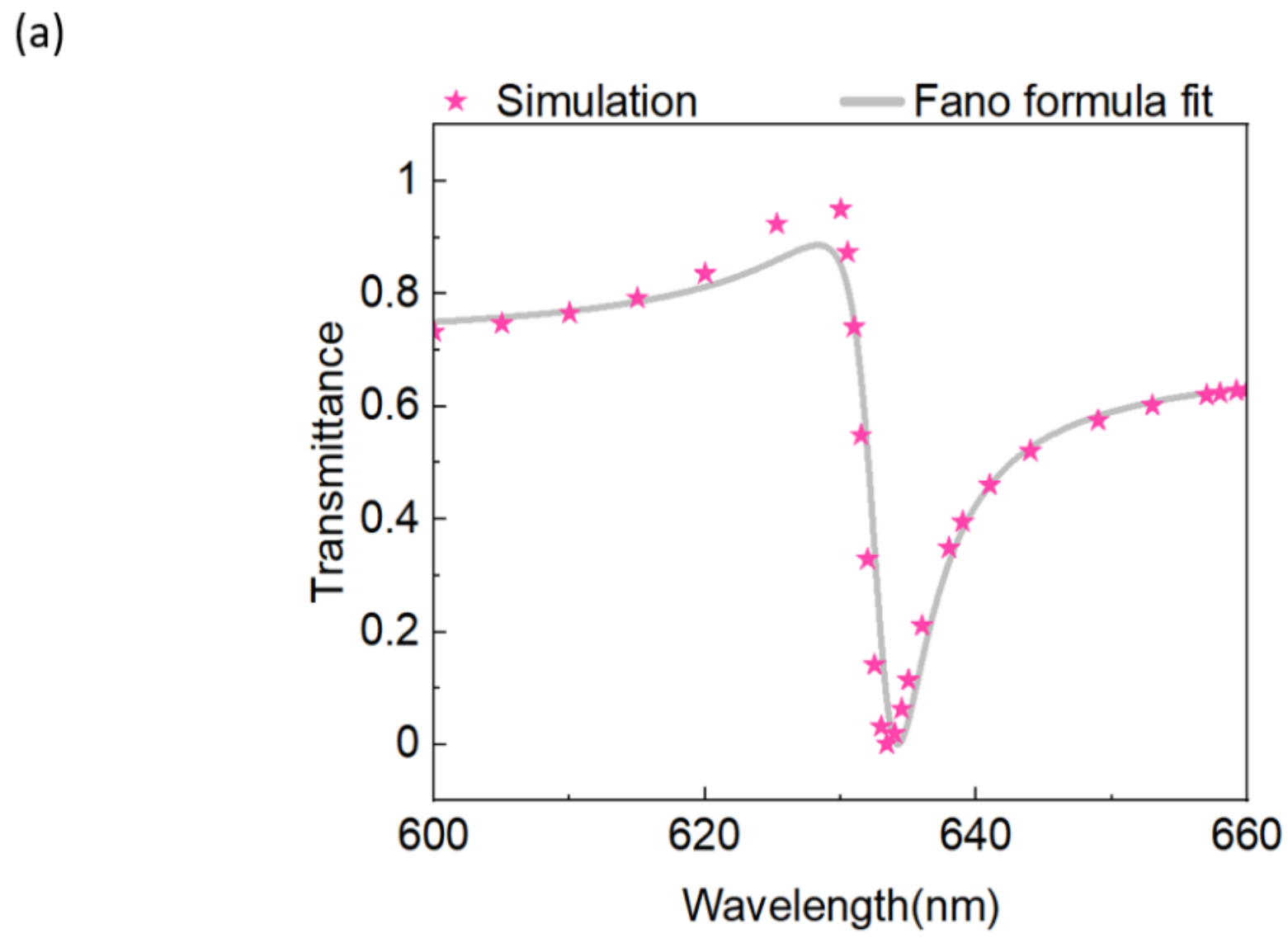

**Figure S4.** Transmission spectrum of bulk $WS_2$ metasurface simulated (red star) and fitted using Fano formula (gray solid curve) when $\Delta G$ = 30 nm.

## 5. Effect of the length of bulk $WS_2$ nanorods on BZF-QBIC mode.

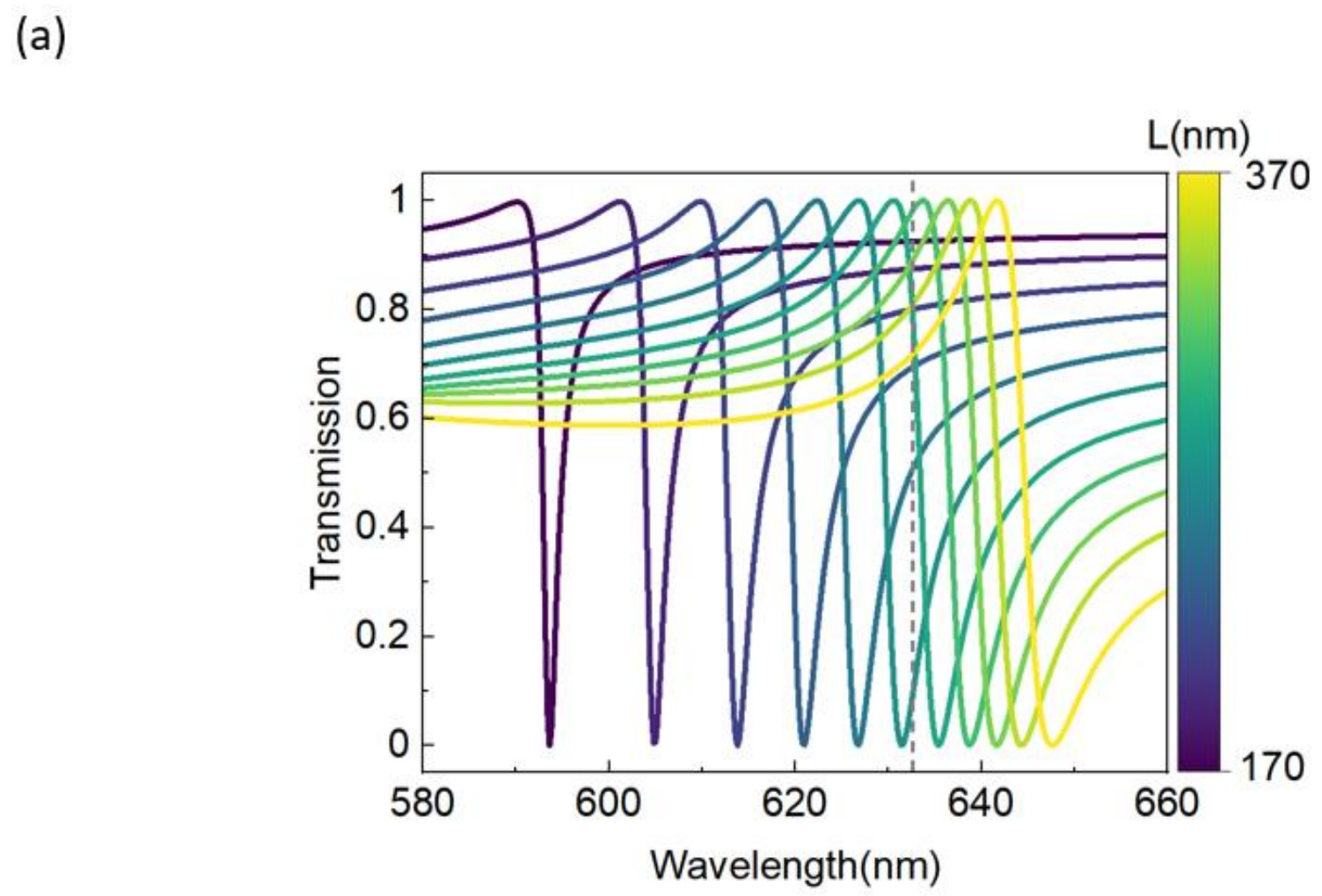

**Figure S5.** Transmission spectrum of bulk $WS_2$ metasurfaces as a function of rod length $L$ with $\Delta G$ = 30 nm $H$ = 70 nm and $W$ = 100 nm. The white dashed lines represent the the bulk $WS_2$ A-exciton wavelength.

## 6. Effect of the length of bulk $WS_2$ nanorods on strong coupling.

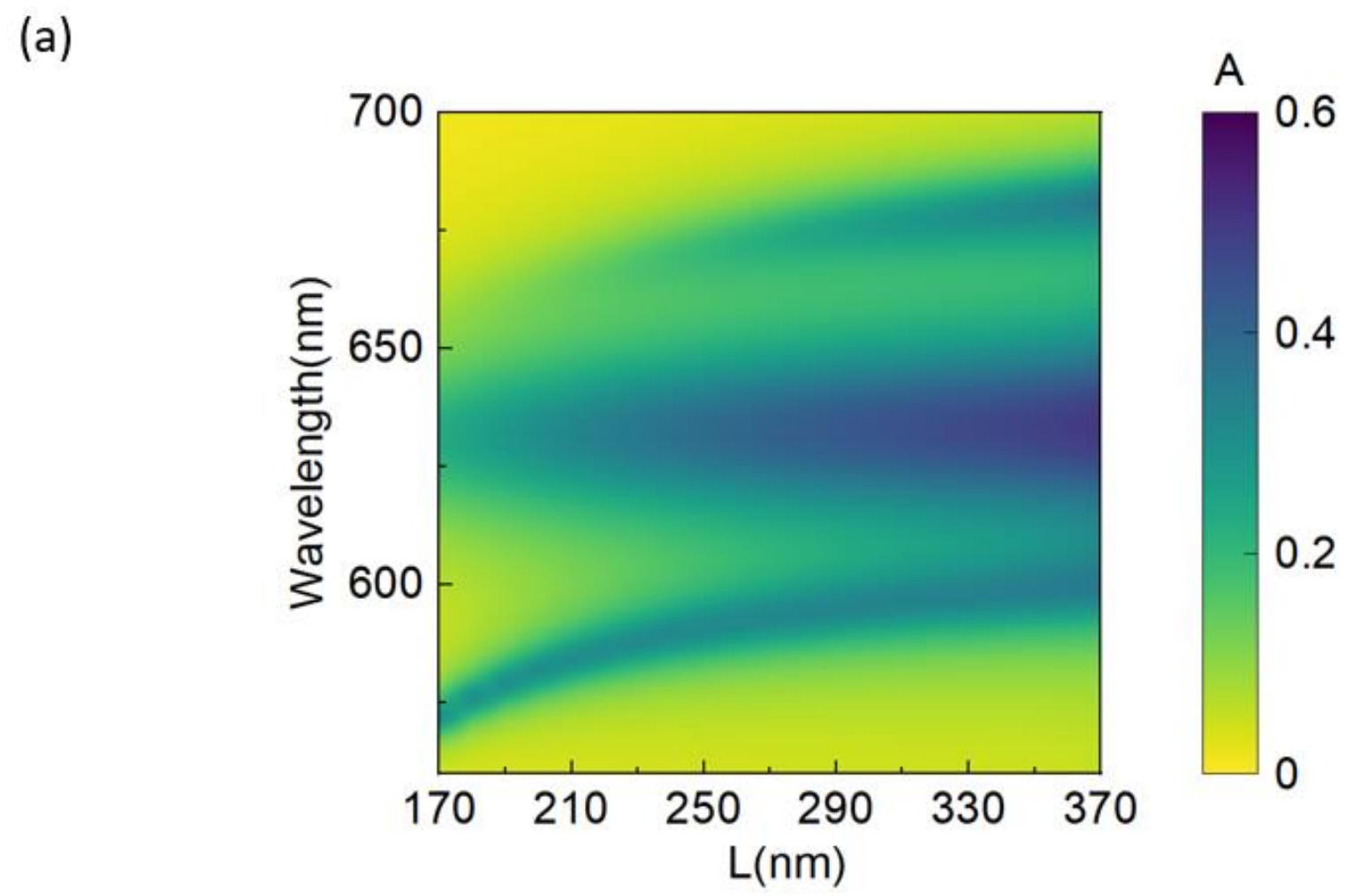

**Figure S6.** The absorption spectrum of the self-hybridized structure as a function of $WS_2$ nanorods length $L$ with the gap perturbation $\Delta$G=30nm.

## 7. Comparison between FEM calculations and CMT model.

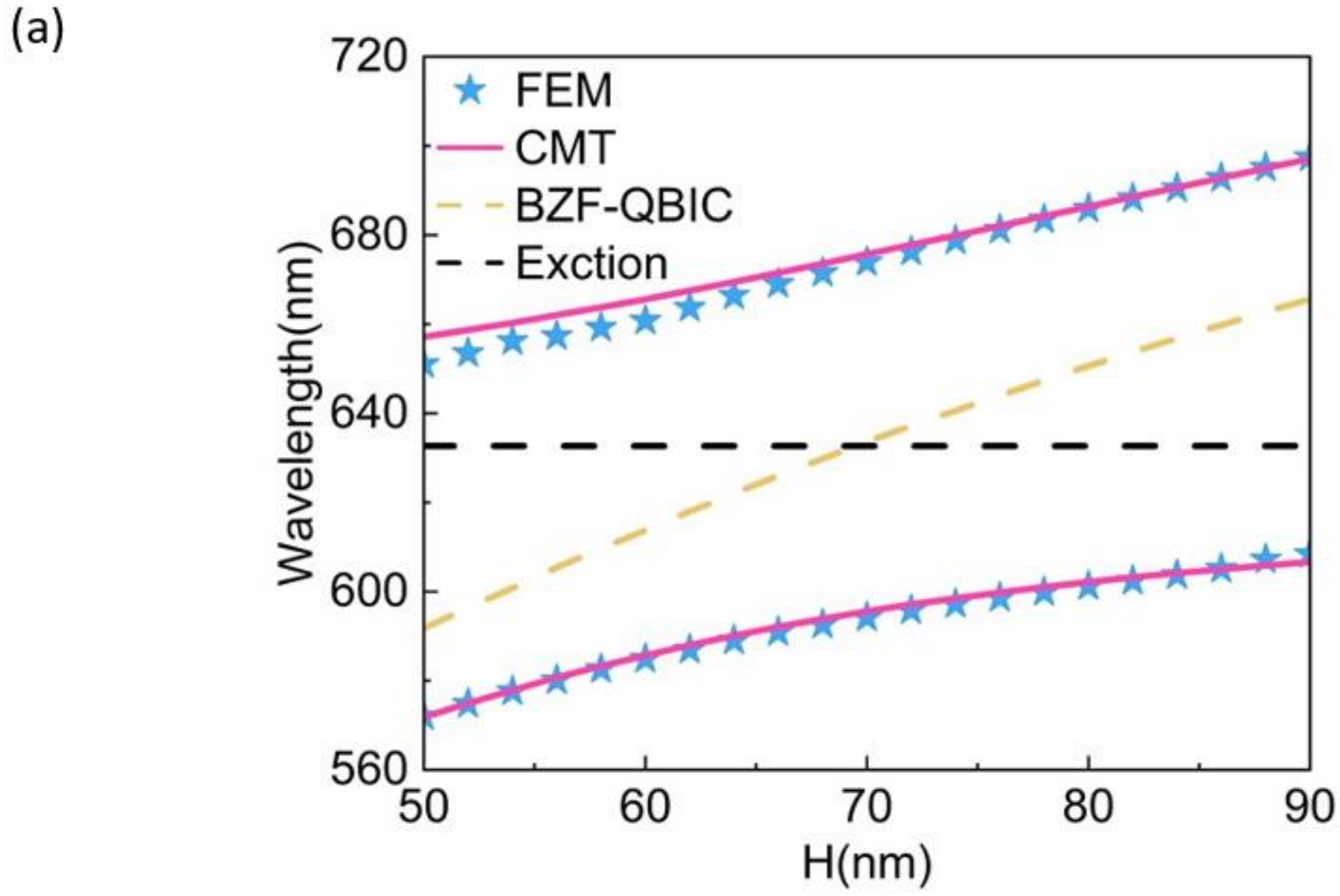

**Figure S7.** (a) The dispersion relations of the coupled system are obtained by FEM calculations (blue stars) and the CMT model (red solid curves). The yellow and black dashed lines represent the dispersion of the uncoupled BZF-QBIC mode and *A* excitons, respectively.

## 8. Height-dependent absorption spectra with different gap perturbations.

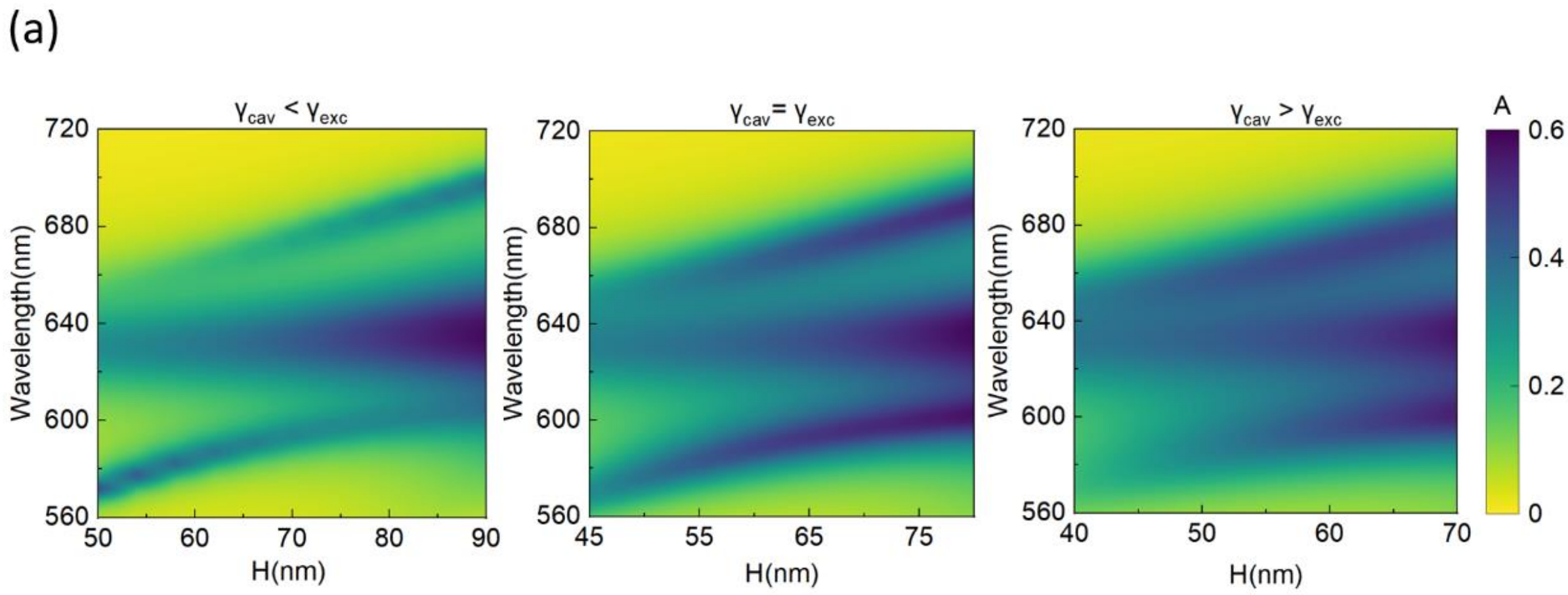

**Figure S8.** The absorption spectra of the coupled structure as a function of $WS_2$ nanorods height *H* with different gap perturbations.